\documentclass{basi}
\usepackage{graphicx}
\begin{document}

\title{LIGHT CURVE SOLUTIONS OF ECLIPSING BINARIES IN SMC}

\author[Valentin P. Ivanov, Diana P. Kjurkchieva and M. Srinivasa Rao]%
       {V. Ivanov $^{1}$\thanks {email: viva.nov@abv.bg},
       D. Kjurkchieva$^{1}$\thanks {email: d.kyurkchieva@shu-bg.net}
and
M. Srinivasa Rao$^{2}$\thanks{email: msrao@iiap.res.in}\\
$^{1}$ Department of Physics, Shumen University, 9700 Shumen, Bulgaria \\
$^{2}$ Indian Institute of Astrophysics, Bangalore 560034, India}
%\pubyear{2001}
%\volume{29}
%\setcounter{page}{17}
%\date{Received 2001 May 30; accepted 2001 June 07}

\maketitle

\begin{abstract}
We propose a procedure for light-curve solution of eclipsing
binary stars in the Small Magellanic Cloud for which photometric
data have been obtained in the framework of the OGLE project as
well as way of determination of the global stellar parameters on
the basis of the obtained solutions, some empirical relations as
well as the distance to the SMC. Several examples illustrate this
procedure.
\end{abstract}

\begin{keywords}
eclipsing binaries; light curves; modeling; Small Magellanic Cloud;
global parameters
\end{keywords}

\section{Introduction}

The microlensing experiments monitor very crowded fields and after
several years they lead to recording of variability of a huge
number of objects and epochs. As a result, the ability to generate
data far exceeds the ability to process it. The microlensing
project OGLE (Optical Gravitational Lensing Experiment) as well as
the projects EROS (Grison et al. 1995) and MACHO (Alcock et al.
1997) monitored millions of stars in the Magellanic Clouds during
the last two decades. This large photometric database is available
not only for microlensing studies but also for individual and
statistical investigations of variable stars.

The OGLE data have been obtained by 1.3-m Warsaw telescope at Las
Campanas Observatory (Chile) equipped with CCD mosaic camera
8kMOSAIC consisting of eight thin SITe 2048 x 4096 CCD chips. The
majority of observations are made in $I$ color. The main results
from the analysis of the OGLE observations are detections of:
gravitational lensing events; planetary and low luminosity object
transits; small amplitude variable red giants in the Magellanic
Clouds; eclipsing stars in the SMC; RR Lyr stars in the LMC;
population II Cepheids in the Galactic Bulge (Paczynski et al.
1999); stellar proper motion; star clusters, etc. In particular,
the OGLE II project (Udalski et al. 1997) provided $BVI$ light
curves for about several millions of stars from the central parts
of the Magellanic Clouds (Udalski et al. 1998, 2000).

The sample of 68000 variable stars detected in LMC and SMC (Graczyk
2003) is reasonably complete allowing statistical analysis and
provides a good material for testing the evolutionary theory of the
binary systems, studying the evolution of the Magellanic Clouds,
star formation, etc. The future spectral observations of the
eclipsing binaries with 8-m European Southern telescopes in Chile
will allow to study the geometrical and spatial distribution of the
stars in these galaxies.

\section{OGLE catalogue of eclipsing binary stars in the SMC}

The search for variable stars in the Magellanic Clouds in the
framework of the OGLE project has been performed automatically
using $I$-band data from 140-180 epochs (with lower limit set of
50) with the software DIA (Difference Image Analysis). The light
curves of the selected candidates are searched for periodicity
using the AoV algorithm (Zebrun et al. 2001).

Udalski et al. (1998) presented the first catalogue of about 1500
eclipsing stars found in the central 2.4 square degree area of the
SMC brighter than $I=20^{m}$ with periods from 0.3 days to 250
days. The next catalogue of variable stars in the Magellanic
Clouds found on OGLE-II data in 1997-2000 (Zebrun et al. 2001)
covers about 7 square degrees of the sky (21 fields in the LMC and
11 fields in the SMC). This catalogue is available in electronic
form: \emph{http://www.astrouw.edu.pl/~ftp/ogle} or
\emph{http://www.astro.princeton.edu/~ogle}.

The OGLE catalogue of the eclipsing binary stars in the SMC
contains 1350 objects of different types. Some of them reveal
eccentric orbits with possible apsidal motion. The binaries are
sorted by type (EA, EB, EW, approximately corresponding to
detached, semi-detached and contact binaries) as well as by the
field they are located in. The catalogue contains the following
information: number of the star; OGLE-II field; name of the star;
RA; DEC; x and y coordinates in the reference image; orbital
period; HJD for the primary maximum (T - 2450000); zero epoch
(phase 0) corresponding to the deeper eclipse; $I$-band brightness
at maximum; $I$-band amplitude (depth of the primary minimum);
phase of the secondary eclipse; depth of the secondary eclipse in
$I$ band; $I$, $B$ and $V$ brightness at maximum; $(B-V)$ and
$(V-I)$ color indices; type of the eclipsing star. Clicking the
star name provides its photometry in the following format: HJD,
magnitude and magnitude error. The errors of magnitude
measurements are: 0.005$^{m}$ for the brightest stars $(I <
$15$^{m})$; 0.08$^{m}$ for stars with 15$^{m} < I < $19$^{m}$;
0.3$^{m}$ for stars with 20.5$^{m} < I$. Typically, there are
about 400 data points in $I$ band and about 30-40 points in $V$
and $B$ band for each variable in the catalogue.

\section{Procedure of light curve solution}

The light curve solution presents searching for the best coincidence
between a family of theoretical light curves corresponding to
different values of the star parameters with observed photometric
data. It is based on the least squares method and the searched
solution is that for which the sum of the residuals is minimum.

Taking into account the quality of the OGLE data we propose a
procedure of light curve solution for the eclipsing binaries in
SMC containing four stages:

(a) preliminary light curve solution on the basis of known
statistical relations between the global stellar parameters
(Kjurkchieva et al. 2006, 2007, 2008);

(b) initial light curve solution by the code Binary Maker 3
(Bradstreet $\&$ Steelman 2004) for determination of relative
stellar radii $r_{1}$ and $r_{2}$, temperatures $T_{1}$ and $T_{2}$,
mass ratio $q$, orbital inclination $i$, eccentricity $e$ and
(eventually) spot parameters;

(c) final multicolor solution using the code DC (Wilson $\&$ Van
Hamme 2003) or PHOEBE (Prsa $\&$ Zwitter 2005) to determine the
varied parameters and their errors;

(d) calculation of the global stellar parameters on the basis of
some empirical relations as well as the distance to the binary, i.e.
to the SMC (Graczyk 2003).

\subsection{Preliminary solution}

The goal of this stage is to get some preliminary values of the
stellar parameters.

The reducing of the catalogue data includes the following steps:

(a) creating the .dat file (phase/flux) from the observed data in
the form JD/magnitude (the value 1 of the flux should correspond
to the middle level at the phase of maximum brightness in $I$
color);

(b) visualization of the .dat file and removing the points that
are far outside the common course of light variability;

(c) measurement of the fluxes $l_1$ and $l_2$ at the bottoms of the
two $I$ minima;

(d) visual estimation of the maximum observed out-of-eclipse flux in
$I$, $B$ and $V$ colors;

(e) determination of the $(B-V)_0$ index by the formula (Harries et
al. 2003)
\begin {equation}
(B-V)_{0}=(B-V)-E_{B-V}=(B-V)-A_V/R
\end{equation}
where $R$=3.1 is the mean total extinction to SMC (Bouchet et al.
1985) and $A_{V}$ is the local extinction to the region surrounding
the target binary determined by 30 stars (Zaritsky et.al 2002). The
values of $A_{V}$ could be taken from {\it SMC Extinction Retrieval
Service}: {\it http://ngala.as.arizona.edu/dennis/smcext.html}.

Then the determination of the preliminary stellar parameters goes in
the following order:

(a) determination of the mean temperature of the binary $T_{0}$ by
the empirical relation ${T_{0}}/{(B-V)_0}$ built on the
calibration of Flower (1996) for the Galactic stars;

(b) calculation of the initial temperature of the secondary star
$T_{02}$ by the formula (Bronstein 1972):
\begin {equation}
T_{02} \approx T_{01} \bigg( {\frac {1-l_2} {1-l_1} }\bigg) ^{1/4}
\end{equation}
adopting that the initial primary-star's temperature $T_{01}$ is
equal to the mean temperature of the binary $T_{0}$, i.e.
$T_{01}$=$T_{0}$;

(c) calculation of the initial mass ratio $q_{0}$ by the empirical
relation (Kjurkchieva $\&$ Ivanov 2006):
\begin{equation}
q_{0} = M_2/M_1 \approx (T_{02}/T_{01})^{1.7};
\end{equation}

(d) calculation of the ratio of relative radii
by the empirical relation (Kjurkchieva $\&$ Ivanov 2006):
\begin{equation}
r_{2}/r_{1}=R_{2}/R_{1} \approx q_{0}^{0.75}.
\end{equation}

\subsection{Initial light curve solution}

In order to get fast, initial, light-curve solution we propose to
use the code Binary Maker 3 (BM3) because it allows to see
immediately the effect of changing of each parameter on the
synthetic light curve (Bradstreet \& Steelman 2004). The
visualization is very useful tool at this stage of trials and
errors.

To obtain synthetic light curve by BM3 one should insert: (a) the
preliminary values of the star parameters obtained by the
empirical relations (stage 1); (b) tabular values of the limb
darkening, gravitation darkening and reflection coefficients
appropriate to the initial temperatures of the stellar components;
(c) some suspected value of the orbital inclination.

To get better coincidence of the theoretical light curve with the
photometric points one should begin to vary the configuration
parameters by trials and errors. Besides the visual estimation of
the fit quality the code BM3 provides the sum of the O-C residuals
as an objective criterion.

We note that: (a) at this stage good fit is searched only in $I$
color because there are a few $B$ and $V$ data of the OGLE
catalogue; (b) the primary's temperature remains fixed at this
stage of procedure; (c) if the observed light curve is asymmetric
and/or distorted one should try solution with surface spots
varying their parameters.

Our experience in the light curve analysis leads to the following
recommendations: (a) The wanted widths of the eclipses may be
reached varying the relative radii; (b) Simultaneous
increase/decrease of the depths of the two minima may be provided
by increase/decrease of the orbital inclination; (c) The
increase/decrease of the relative depth of the secondary minimum
can be provided by increase/decrease of the temperature of the
secondary star; (d) Spots with longitudes around $90 ^\circ$ and
$270^\circ$ change the out-of-eclipse light levels while spots
with longitudes around $0^\circ$ and $180^\circ$ change the shape
and depth of the corresponding eclipse. As a rule, the longitude
of the spot center corresponds to the maximum of the light curve
distortion.

If the foregoing procedure leads to solution with $q > 1$ then the
parameters of the two stars (temperatures, darkening coefficients,
albedoes and relative radii) should be exchanged $by$ $hand$. This
is necessary because we assume that the primary star is eclipsed
during the deeper minimum while BM3 assumes as a primary star that
with bigger mass.

After reaching a good $I$-band solution one should obtain (by BM3)
the corresponding $V$ and $B$ synthetic curves. Usually they
reproduce well the observational points (when the corresponding
stars emit as black body). Moreover, the good coincidence in $V$
and $B$ bands means that we have determined well the maximum light
in $B$ and $V$ colors. If this is not the case, we should search
for a better coincidence of the light curves in all three colors
simultaneously varying the normalization level of the respective
curve ($B$ or/and $V$) on the basis of visual estimation. This
leads to some correction of the color index as well as the primary
temperature and consequently to repeating of the whole procedure
but varying the parameters in quite narrow ranges.

Due to the slow calculation of eccentric orbits by BM3 we
recommend to begin those cases with a circular-orbit solution
aiming coincidence of the depths and widths of the eclipses. After
that, one should input suitable values of the eccentricity and the
argument of periastron (the OGLE catalogue gives some initial
values for them) and continue the fitting.

After reaching a good fit by BM3, one could use the obtained initial
stellar temperatures $T_{01}$, $T_{02}$ and the ratio of the
relative luminosities $k$ = $l_{2}$/$l_{1}$ for disentangling the
temperatures $T_{1}$ and $T_{2}$ by the Rayleigh-Jeans approximation
(the luminosity to be a linear function of temperature) by the
formulae (Graczyk 2003):
\begin{equation}
T_{1}=T_{0}(1+k)/(1+kT_{02}/T_{01})
\end{equation}
\begin{equation}
T_{2}= T_{1}(T_{02}/T_{01}).
\end{equation}

\subsection{Final light curve solution}

The code LC (similarly to BM3) calculates synthetic light curves on
the basis of black body emission and certain parameters of the
binary. It works in three different modes: detached, semidetached
and contact configuration. However, the LC code has not a
possibility for visualization and fast estimation of the fit
quality. That is why in our procedure we use the code LC only for
creation a table with input parameters in the format appropriate for
the code DC (Wilson 1992, Wilson $\&$ Van Hamme 2003). For this aim
we run the code LC once with the values obtained by the code BM3.

The code DC uses the method of differential corrections for a
least squares analysis (Wilson $\&$ Devinney 1971). DC needs table
of input parameters from LC as well as photometric data. The user
could choose the adjusted parameters. It is recommended the step
of their varying to be around 1 $\%$ of the input value of the
corresponding parameter, for instance: 100 K for the temperature
of the secondary star $T_2$; $0.1^\circ$ for the orbital
inclination $i$; 0.005 for the relative star radii $r_i$; 0.01 for
the mass ratio $q$; $5^\circ$ for the spot latitude $\beta_{sp}$
and spot longitude $\lambda _{sp}$; $1^\circ$ for the angular spot
radius $\alpha_{sp}$; 0.01 for the relative spot temperature
$T_{sp}/T_{st}$.

The DC code makes iterations and presents the results in output
file. It contains the final values of the adjusted parameters,
their errors and the sum of the residuals for the three light
curves corresponding to their $best$ solution. The number of
iterations depends on the fit quality of the initial solution.

The final solution might be obtained also by the new code PHOEBE
(Prsa $\&$ Zwitter 2005) that incorporates all features of DC but it
has possibilities for visualization of the procedure.

\subsection{Determination of the global stellar parameters}

It is impossible to calculate the absolute dimensions and masses of
binary stars without their spectroscopic orbit but there is a way to
estimate them using the method of the parallaxes (Dvorak 1974).
Graczyk (2003) used this method for the independent estimate of the
interstellar reddening $E_{(B-V)}$ in the direction to a particular
binary.

We propose determination of the global parameters of the eclipsing
binaries on the basis of their light curve solutions and the known
value of the distance to the SMC by the following procedure
(Kjurkchieva $\&$ Ivanov 2008):

(1) The primary's and secondary's visual magnitudes $V_{1}$ and
$V_{2}$ are calculated by the Pogson's formulae
\begin{equation}
V-V_{1}=-2.5 \log(1+k)
\end{equation}
\begin{equation}
V-V_{2}=-2.5 \log(1+1/k)
\end{equation}
where $V$ is the visual magnitude of the binary and $k$ is the ratio
of the relative luminosities that is obtained from the light curve
solution.

(2) The absolute magnitudes $M_{i}^{V}$ of the components can be
calculated by the distance-modulus relation using $DM=18.9^{m}$ for
SMC (Graczyk 2003) and taking into account the interstellar
reddening (Zaritsky et al. 2000)
\begin{equation}
M_{i}^{V}=V_{i}-DM-RE_{(B-V)}  .
\end{equation}
Then one can calculate the bolometric stellar magnitudes
$M_{i}^{bol}$ by the equation
\begin{equation}
M_{i}^{bol}=M_{i}^{V}+BC_{i}
\end{equation}
where the bolometric corrections $BC_{i}$ correspond to obtained
stellar temperatures and they can be taken from Flower (1996).

(3) The absolute luminosities $L_{i}$ can be calculated by the
obtained absolute magnitudes $M_{i}^{bol}$:
\begin{equation}
L_{i}=10^{(4.75-M_{i}^{bol})/2.5}.
\end{equation}

(4) Inserting the last expression as well as $R_{i}=ar_{i}$ into the
formula $L_{i}=4\pi R_{i}^{2}\sigma T_{i}^{4}$ one can obtain the
orbital separation $a$ of the binary
\begin{equation}
\log a=0.2(M^{bol}-M_{i}^{bol}) - \log r_{i} - 2 \log T_{i}+7.524  .
\end{equation}

(5) The absolute stellar radii $R_{i}$ can be obtained from the
expression $R_{i}$=$a r_{i}$ where $r_{i}$ are the relative
stellar radii from the light curve solutions.

(6) The mass of each binary's component can be calculated by the
empirical relation mass-luminosity for MS stars
\begin{equation}
{\cal M}_i=10^{(\log L_{i}^{bol}-0.38)/3.664}.
\end{equation}

(7) The densities $\rho_{i}$ and surface gravity $g_{i}$ of the
individual components could be estimated by the formulae:

\begin{equation}
g_{i}=G{\cal M}_i/R_{i}^{2}
\end{equation}
\begin{equation}
\rho_{i}=3{\cal M}_i/{4{\pi}R_{i}^{3}}
\end{equation}

(8) Finally the spectral type and the luminosity class of each stellar
component can be determined from the values of $T_{i}$, $\rho_{i}$ and
$g_{i}$.

\section{Illustration of the proposed method}

In order to illustrate our method for the light curve solution of
the OGLE data of eclipsing binaries we chose three stars with
circular orbits of different types: detached D, semi-detached Sd
and contact C. Table 1 presents the obtained results from the
light curve modeling and Figure 1 illustrates the fit quality of
their light curve solutions. The first five columns of the table
show respectively the star number, temperatures of both components
$T_{i}$, mean relative stellar radii $r_{i}$, photometric mass
ratio $q$ and orbital inclination $i$. The next three columns
reveal global star parameters in solar units: luminosities
$L_{i}$, absolute radii $R_{i}$ and masses $\cal M_{i}$ while the
last two columns presents the star configuration and spectral type
of the components.

Figure 2 presents the Herzschprung-Russell diagram
for our three stars (marked by asterisks). The figure contains also 
the positions of SMC stars investigated by
Graczyk (2003) and marked by circles as well as those
of Hilditch et al. (2005) marked by triangles. It is visible that
our stars hotter than 15000 K are slightly above ZAMS and their locations
coincide with those from previous studies while the
stars with T$ < $15000 K lie farther from ZAMS.

\begin{table*}
 \begin{center}
  \caption[]{Global parameters of three eclipsing binary stars of our sample}
\begin{tabular}{|c|r|r|r|r|r|r|r|r|r|r|r|}
\hline
Star number&$T_{i}$&$r_{i}$&$q$  & $i$&$L_{i}$&$R_{i}$&$\cal M$$_{i}$&$Conf.$& Sp   \\
\hline
003831.81- & 20900& 0.255 & 0.9  &70.4& 1863 & 3.30 & 6.03 & D    & B3V      \\
 733308.7  &17500 & 0.267 &      &    & 1030 & 3.46 & 7.10 &      & B4V     \\
004444.36-7& 12400 & 0.364 & 0.51 &58.5& 1107 & 7.23 & 5.23 & Sd &
B7V       \\
 31051.5   & 9700 & 0.321 &      &    & 271  & 6.38 & 5.52 &      & A0V     \\
003835.24- & 5400 & 0.413 & 0.76 &64.1& 1065 & 37.41& 5.17 & C    & G2III       \\
 735413.2  & 5100 & 0.365 &      &    & 695  & 33.05& 6.63 &      & G4III     \\
\hline
\end{tabular}
\end{center}
\end{table*}

\begin{figure}
  \begin{center}
% \vspace{1cm}
 \includegraphics[width=13cm,height=5cm,scale=1.00]{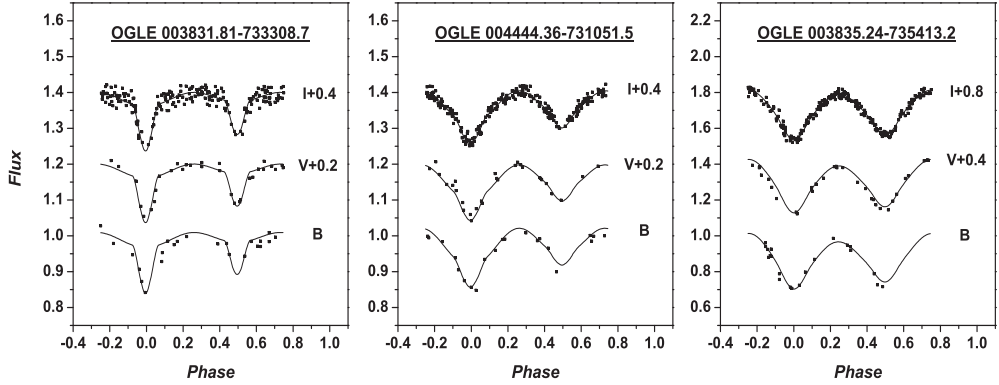}
 \caption[]{Illustrations of the fit quality of our light curve solutions}
 \label{fig1}
 \end{center}
 \end{figure}

\begin{figure}
  \begin{center}
% \vspace{1cm}
 \includegraphics[width=6cm,height=5cm,scale=1.00]{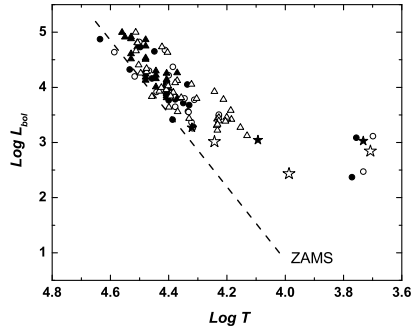}
 \caption[]{Herzschprung-Russell diagram of SMC stars (the filled and empty symbols refer to the primary and secondary components of the eclipsing binaries, the dotted line presents ZAMS)}
 \label{fig2}
 \end{center}
 \end{figure}

\section{Conclusion}

The eclipsing stars are important sources of information
concerning fundamental problems of stellar astrophysics because
they allow determination of the global stellar parameters (radii,
masses, luminosities, stellar composition). Moreover the
investigations of eclipsing binaries in large and homogenous
sample give a possibility to improve the empirical statistical
relations between these parameters and provide empirical tests for
the stellar evolution. This investigation is a small step in the
solution of this important problem.

\textbf{Acknowledgements}. The research was supported partly by
funds of the project DO 02-362 of Bulgarian Science Foundation.

\end{document}